\documentclass[english,aps,manuscript,showpacs,preprint,superscriptaddress]{revtex4-1}%
\usepackage[T1]{fontenc}
\usepackage[latin9]{inputenc}
\usepackage{color}
\usepackage{amsmath}
\usepackage{amssymb}
\usepackage{esint}
\usepackage{amsmath}
\usepackage{latexsym}
\usepackage{amssymb}
\usepackage{epsfig}
\usepackage{amsthm}
\usepackage{epstopdf}
\usepackage{graphicx,subfigure}
\usepackage{color,xcolor}
\theoremstyle{plain}\newtheorem{theorem}{Theorem}[section]
\theoremstyle{definition}\newtheorem{Defi}[theorem]{Definition}
\theoremstyle{plain}
\theoremstyle{remark}
\theoremstyle{plain}

\newcommand{\pa}{\partial}

\def\bi{\begin{align}}
\def\bin{\begin{align*}}
 \def\epa{\end{pmatrix}}
\def\bpa{\begin{pmatrix}}
\newcommand{\ein}{\end{align*}}
\newcommand{\ei}{\end{align}}

\newcommand{\vp}{\varphi}

\newcommand{\sumli}{\sum\limits}
\newcommand{\ba}{\begin{array}}
\newcommand{\ea}{\end{array}}

\makeatletter

\@ifundefined{textcolor}{}
{%
 \definecolor{BLACK}{gray}{0}
 \definecolor{WHITE}{gray}{1}
 \definecolor{RED}{rgb}{1,0,0}
 \definecolor{GREEN}{rgb}{0,1,0}
 \definecolor{BLUE}{rgb}{0,0,1}
 \definecolor{CYAN}{cmyk}{1,0,0,0}
 \definecolor{MAGENTA}{cmyk}{0,1,0,0}
 \definecolor{YELLOW}{cmyk}{0,0,1,0}
}

\usepackage{babel}

\makeatother

\usepackage{babel}
\begin{document}

\title{Explicit non-canonical symplectic algorithms for charged particle dynamics}

\author{Yang He}

\affiliation{School of Nuclear Science and Technology and Department of Modern
Physics, University of Science and Technology of China, Hefei, Anhui
230026, China}

\address{Key Laboratory of Geospace Environment, CAS, Hefei, Anhui 230026,
China}

\author{Yajuan Sun}

\affiliation{LSEC, Academy of Mathematics and Systems Science, Chinese Academy
of Sciences, P.O.Box 2719, Beijing 100190, China}

\author{Zhaoqi Zhou}

\affiliation{LSEC, Academy of Mathematics and Systems Science, Chinese Academy
of Sciences, P.O.Box 2719, Beijing 100190, China}

\author{Jian Liu}

\affiliation{School of Nuclear Science and Technology and Department of Modern
Physics, University of Science and Technology of China, Hefei, Anhui
230026, China}

\address{Key Laboratory of Geospace Environment, CAS, Hefei, Anhui 230026,
China}

\author{Hong Qin}

\affiliation{School of Nuclear Science and Technology and Department of Modern
Physics, University of Science and Technology of China, Hefei, Anhui
230026, China}

\affiliation{Plasma Physics Laboratory, Princeton University, Princeton, NJ 0854}

\begin{abstract}
We study the non-canonical symplectic structure, or $K$-symplectic structure inherited by the charged particle dynamics.
Based on the  splitting technique, we  construct non-canonical symplectic methods which is explicit and stable  for the long-term simulation. The key point of splitting is to decompose the Hamiltonian as four parts, so that the resulting four subsystems have the same structure and can be solved exactly. This guarantees the $K$-symplectic preservation of the numerical methods constructed by composing  the  exact solutions  of the subsystems.
The error convergency  of numerical solutions is analyzed  by means of the Darboux transformation. 
The numerical experiment  display the long-term stability and efficiency for  these  methods.
\end{abstract}

\maketitle
\section{Introduction}

In the application of magnetized plasmas, the motion of charged particles under the influence of  electromagnetic fields is the most fundamental process.  The long-term simulation of the dynamics with better accuracy is significant for various multi-timescale problems, such as particles in tokamaks.  It is a recent successful development to apply the idea of geometric numerical methods \cite{hairerlw06gni}  in   simulating  the dynamics of charged particle equations. In \cite{heslq15vpa, heslq15hov}, according to the volume-preserving nature of the charged particle system the (high-order) numerical methods have been constructed in order to preserve the volume in phase space. Compared with the standard  methods such as the fourth order Runge-Kutta method, this kind of volume-preserving numerical methods  can bound the errors of  conservative laws of the given system, and yield accurate numerical result over long time. 

In a given electromagnetic field ${\bf E(x)}$ and ${\bf B(x)}$, the motion of a single particle admits  the Lorentz force law  with which the dynamical system  can be described as
\begin{align}\label{eq:lfl}
\dot{\bf x}={\bf v}, \quad
m\dot {\bf v}=q\left({\bf E(x)}+{\bf v}\times {\bf B(x)}\right).
\end{align}
It is well known that the system has a  canonical structure in coordinates ${(\bf x}, {\bf p})$\cite{heslq15hov}. However, generally classical canonical symplectic methods for this system can not be implemented explicitly.
 Notice that Eqs.~(\ref{eq:lfl}) itself is separable and equipped with a non-canonical symplectic structure. It motivates us to construct  a new kind of explicit numerical methods based on this property.

With this purpose, we rewrite (\ref{eq:lfl}) as
\begin{equation}\label{eq:lorentz}
\left(\begin{array}{cc}
-q\hat{\bf B}({\bf x})& -mI\\
mI & 0
\end{array}\right)\left(
  \begin{array}{c}
 \dot{{\bf x}}  \\
   \dot{{\bf v}}  \\
  \end{array}
   \right)
  =
   \left(
            \begin{array}{c}
              q\nabla \vp({\bf x}) \\
              m {\bf v} \\
            \end{array}
   \right),
\end{equation}
where $
{\hat{\bf B}({\bf x})}=\left(\begin{smallmatrix}
0 & -B_{3}({\bf x}) & B_{2}({\bf x})\\
B_{3}({\bf x}) & 0 & -B_{1}({\bf x})\\
-B_{2}({\bf x}) & B_{1}({\bf x}) & 0
\end{smallmatrix}\right)$ is defined by the magnetic field ${\bf B}$. It is in a more compact form as
\begin{equation}\label{ksym}
\dot{\bf z}=K^{-1}({\bf z})\nabla H({\bf z}),
\end{equation}
with ${\bf z}=[{\bf x}^\top,{\bf v}^\top]^\top$, $K({\mathbf z})$ the skew-symmetric matrix on the left side of Eq. (\ref{eq:lfl}),
 and   $H({\bf x,v})=m{\bf v}\cdot{\bf v}/2+q\vp({\bf x})$ the energy.  Denote  a two-form in  phase space $({\bf x}, {\bf v})$ with the entries of $K({\bf z})$ as
\begin{equation}\label{eq:2form}
\Omega=q/2 {\rm d}{\bf x}\wedge \hat{\bf B}({\bf x}){\rm d}{\bf x}+m {\rm d}{\bf x}\wedge \rm{d}{\bf v}.
\end{equation}
It is clear from (\ref{eq:2form}) that  $\rm{d}\Omega=0$ for $\nabla \cdot {\bf B}({\bf x})=0$. Thus,  the skew-symmetric matrix $K({\bf z})$  defines  a non-canonical symplectic
structure in $({\bf x}, {\bf v})$. This implies that the phase flow $\vp_t: {\bf z}\mapsto\vp_t({\bf z})$ of system (\ref{ksym})  preserves the $K$-symplectic structure, which is
\begin{equation}\label{eq:ksystr}
\frac{d}{dt}{\rm d}\vp_t({\bf z})\wedge K(\vp_t({\bf z})){\rm d}\vp_t({\bf z})=0.
\end{equation}

The purpose of the current work is to present  a class of explicit $K$-symplectic structure-preserving methods.  They are constructed   using the Hamiltonian splitting which decomposes  the given system as several solvable subsystems.
 We organize the paper as follows.  In section 2, we present the construction of $K$-symplectic numerical methods based on splitting technique. We also  provide the theoretical results on error of numerical solutions by Darboux transformation.  Experiments are shown in section 3. Section 4 concludes this paper.
For simplicity, in the following discussions the equations are all normalized.

\section{$K$-symplectic structure-preserving methods}
We start this section by   introducing  a  definition of $K$-symplectic structure-preserving methods.
\begin{Defi}\label{de:1}
A numerical method ${\bf z}_{n+1}=\vp_h({\bf z}_n)$  applied to system (\ref{ksym})   is called $K$-symplectic structure-preserving
if it satisifes
\begin{equation}\label{eq:disksy}
{\rm d}\vp_h({\bf z}_n)\wedge K(\vp_h({\bf z}_n)){\rm d} \vp_h({\bf z}_n)={\rm d} {\bf z}_n \wedge K({\bf z}_n){\rm d}{\bf z}_n,
\end{equation}
or, alternatively
\begin{equation}\label{eq:disksy1}
\left(\frac{\pa \vp_h}{\pa {\bf z}_n}\right)^\top K(\vp_h({\bf z}_n))\frac{\pa \vp_h}{\pa {\bf z}_n}=K({\bf z}_n).
\end{equation}
\end{Defi}
It is clear that (\ref{eq:disksy}) is a discretization of (\ref{eq:ksystr}). The questions arises as to how the $K$-symplectic-preserving numerical methods can be constructed. Already it is known that the traditional  numerical methods can not, in general,  preserve the $K$-symplectic structure.  Nevertheless, possible construction can be accomplished if  the concerned  system can be split as several solvable subsystems which share   the same structure as the original system.
Due to this,  we rewrite the Hamiltonian as
$$H({\bf x,v})=\frac{1}{2}\sumli_{i=1}^3v_i^2+\vp({\bf x}):=\sumli_{i=1}^3H_{v_{i}}+H_\vp.$$
The Lorenz force system \label{eq:lorentz} is, therefore, equivalent  to
\begin{equation}
{\dot {\bf z}}=\sumli_{i=1}^3K^{-1}({\bf z})\nabla H_{v_i}+K^{-1}({\bf z})\nabla H_\vp.
\end{equation}
This leads to  four subsystems possessing the same structure (\ref{eq:ksystr}). It is observed that all the subsystems can be solved explicitly. The solutions are displayed as below.

Firstly we consider the dynamics generated by $H_{\vp}$, it   is
\begin{equation}
\begin{aligned} & \dot{{\bf x}}={\bf 0},\\
 & \dot{{\bf v}}={\bf E(x)}.
\end{aligned}
\label{eq:subVP}
\end{equation}
It is obviously an integrable system. From the point ${\bf z}(t)=({\bf x}(t),{\bf v}(t))$, the exact solution flow of this subsystem denoted by ${\bf z}(t+h)=\phi_{h}^{H_{\vp}}({\bf z}(t))$ is
\begin{equation}\label{eq:SolHvp}
\left\{\begin{aligned}
{\bf x}(t+h)&={\bf x}(t),\\
{\bf v}(t+h)&={\bf v}(t)+h{\bf E}({\bf x}(t)).
\end{aligned}\right.
\end{equation}

For a given $i$ ($i=1, 2, 3$), the subsystem associated with $H_{v_{i}}$
is
\begin{equation}\label{eq:EqHvi}
\begin{aligned}\dot{{\bf x}}&=v_{i}{\bf e}_{i},\\
\dot{{\bf v}}&=v_{i}\sumli_{j=1}^3{\hat B}_{ij}({\bf x}){\bf e}_j,
\end{aligned}
\end{equation}
where ${\bf e}_i$ is the three-dimensional vector with the $i$-th element being 1, and ${\hat B}_{ij}$ is the entry in the $i$th row and $j$th column of ${\hat {\bf B}}$. As $v_i$ is constant along time, the subsystem is also exactly solvable.
From the given point $({\bf x}(t),{\bf v}(t))$, the solution of the subsystem (\ref{eq:EqHvi}) denoted by ${\bf z}(t+h)=\phi_{h}^{H_{v_{i}}}({\bf z}(t))$ is
\begin{equation}\label{eq:SolHvi}
 \left\{\begin{aligned}
{\bf x}(t+h)&={\bf x}(t)+hv_{i}(t){\bf e}_{i},\\
{\bf v}(t+h)&={\bf v}(t)+\sumli_{l=1}^3 {\bf e}_{l}\int_{x_i(t)}^{x_i(t)+hv_i(t)}{\hat B}_{il}({\bf x}(t))dx_i,\\
\end{aligned}\right.
\end{equation}
More specifically, when $i=1$  the solution  is 
\begin{align*}
{\bf x}(t+h)&={\bf x}(t)+hv_{1}(t){\bf e}_1,\\
{\bf v}(t+h)&={\bf v}(t)+F_{2}^{(1)}({\bf x}(t), {\bf v}(t),h){\bf e}_{2}+F_3^{(1)}({\bf x}(t), {\bf v}(t),h){\bf e}_{3},
\end{align*}
where $
F_2^{(1)}({\bf x}, {\bf v},h)=-\int_{x_1}^{x_{1}+hv_1}B_{3}(\xi,x_{2},x_{3})d\xi
$
and
$F_3^{(1)}({\bf x}, {\bf v},h)=\int_{x_1}^{x_{1}+hv_1}B_{2}(\xi,x_{2},x_{3})d\xi.$
We note that the integrals in Eq. (\ref{eq:SolHvi}) should be evaluated exactly, which is feasible
when the magnetic field is given in terms of familiar functions. When the integrals cannot
be evaluated exactly, we can chose a discretization of the magnetic field such that the
integrals can be exactly evaluated under this discretization. For example, we can discretize
the magnetic field on a Cartesian grid using piece-wise polynomials. Then the non-canonical
symplectic structure associated with a given discretized magnetic field is preserved exactly by $\phi_h^{H_{v_i}}$. In practice, the magnetic field is often specified discretely.

By composing the solutions in Eqs. (\ref{eq:SolHvp}) and (\ref{eq:SolHvi}) of the four subsystems, we  obtain a $K$-symplectic method 
for the Lorenz force system (\ref{ksym})  with the accuracy of order 1  as
\begin{equation}\label{eq:sch1}
\Phi_{h}^1=\phi^{H_{v_{1}}}_h\circ\phi^{H_{v_{2}}}_h\circ\phi^{H_{v_{3}}}_h\circ\phi^{H_{\vp}}_h.
\end{equation}
The $K$-symplectic  method up to second order accuracy can be gained by a   symmetric composition, which is
\begin{align}\label{eq:sch2}
\Phi_{h}^2=\phi^{H_{v_{1}}}_{h/2}\circ\phi^{H_{v_{2}}}_{h/2}\circ\phi^{H_{v_{3}}}_{h/2}\circ\phi^{H_{\vp}}_{h}\circ\phi^{H_{v_{3}}}_{h/2}\circ\phi^{H_{v_{2}}}_{h/2}\circ\phi^{H_{v_{1}}}_{h/2}.
\end{align}
By  definition \ref{de:1}, it is clear that the methods (\ref{eq:sch1}) and (\ref{eq:sch2}) are  $K$-symplectic because of the group property.  Furthermore, it is possible to increase the accuracy of the numerical methods by various compositions \cite{mclachlanq02sm, hairerlw06gni}. It is obvious that the numerical methods   (\ref{eq:sch1}) and (\ref{eq:sch2}) presented here are   all explicit and easy to compute.

Taking the determinate of (\ref{eq:disksy1}) on both sides, we get
 $${\rm det}\left(\frac{\pa \vp_h}{\pa {\bf z}_n}\right)^2{\rm det}(K(\vp_h({\bf z}_n))) ={\rm det}(K({\bf z}_n)).$$
The above equality provides ${\rm det}\left(\frac{\pa \vp_h}{\pa {\bf z}_n}\right)=1$ via  using  the definition of $K(z)$.
This   implies that for the Lorentz force system the $K$-symplectic method is also volume-preserving.

By Darboux's theorem, there exists a coordinate transformation ${\bf z}=T({\bf y})$ such that a non-canonical system (\ref{ksym}) can be turned into
a canonical system
\begin{equation}\label{eq:jsym}
\dot{\bf y}=J^{-1}\nabla {\bar H}({\bf y}),
\end{equation}
where ${\bar H}({\bf y})=H(T({\bf y}))$.  Under the coordinate transformation,  a $K$-symplectic numerical method (\ref{eq:disksy1}) is also  transformed to a symplectic method.
As ${\bar H}({\bf y})=H(T({\bf y}))=H({\bf z})$, under the conditions of Theorem~IV.8.1 in \cite{hairerlw06gni} we have the numerical  energy error estimate for  the $K$-symplectic methods of order $p$
\begin{align}\label{eq:err}
&H({\bf z}_n)=H({\bf z}_0)+{\cal O}(h^p), \quad nh\leq e^{h_0/2h}.
\end{align}
Assume that the transformation function $T$ is bounded,  under the condition of  Theorem~X.3.1 we have the following error estimate for the numerical solution
\begin{align}
||{\bf z}_n-{\bf z}(t)||\leq Cth^p, \quad \hbox{for}\,\, t=nh\leq h^{-p}.
\end{align}
These results guarantee that the $K$-symplectic methods can bound the energy error over exponentially long time. Moreover, the global solution error is restricted to linear growth, which is much smaller than the quadratic or exponential error growth of standard methods.

\section{Numerical experiments}
In this section, we test the non-canonical symplectic methods presented in the above section  by simulating the charged particle dynamics in a given  electromagnetic field.
We consider the two dimensional dynamics of a charged particle under the symmetric static electromagnetic field,
\begin{align}\label{eq:mag}
{\bf B}=R{\bf e}_3, \quad {\bf E}=\frac{10^{-2}}{R^3}(x_1{\bf e}_1+x_2{\bf e}_2)
\end{align}
where $R=\sqrt{x_1^2+x_2^2}.$ 
In the implementation of    the numerical method (\ref{eq:sch1}) or (\ref{eq:sch2}), it is noticed from (\ref{eq:SolHvi}) that we need to calculate an integral of the magnetic field ${\bf B}$. It usually can be computed explicitly. In this example, integrating ${\bf B}$ with respect to $x_1$ gives
\begin{align*}
\int_{x_1-tv_1}^{x_1} B_3(\xi,x_2,x_3) d\xi=f(x_1,x_2)-f(x_1-tv_1,x_2)
\end{align*}
with $f(x_1,x_2)=x_1\sqrt {{x_1}^{2}+{x_2}^{2}}/2+{x_2}^{2}\ln  \left( x_1+\sqrt {{x_1}^{2}
+{x_2}^{2}} \right)/2$.

Firstly, we illustrate  the long-term behavior of  a second order non-canonical symplectic method,  compared with the traditional 4-th order Runge-Kutta method.  The initial position and velocity are taken as ${\bf x}_0=-{\bf e}_2$ and ${\bf v}_0=0.2{\bf e}_1+0.1{\mathbf e}_2$. With the step size $h=\pi/10$, we perform the two methods over the time $[0, 2\times10^4 h]$.
\begin{figure}[t]\centering
\subfigure[]
{\includegraphics[width=6cm,height=5.5cm]{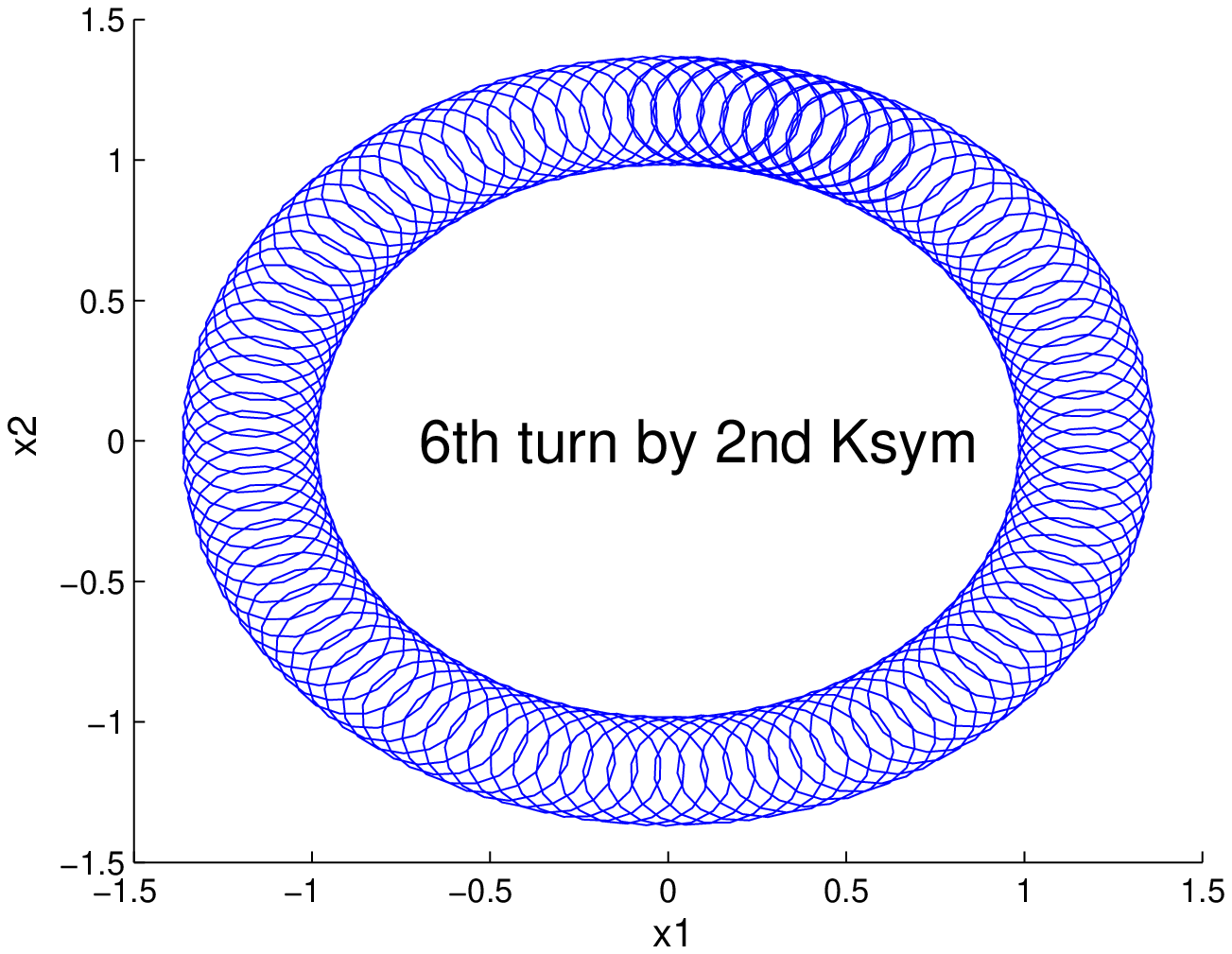}}
\subfigure[]
{\includegraphics[width=6cm,height=5.5cm]{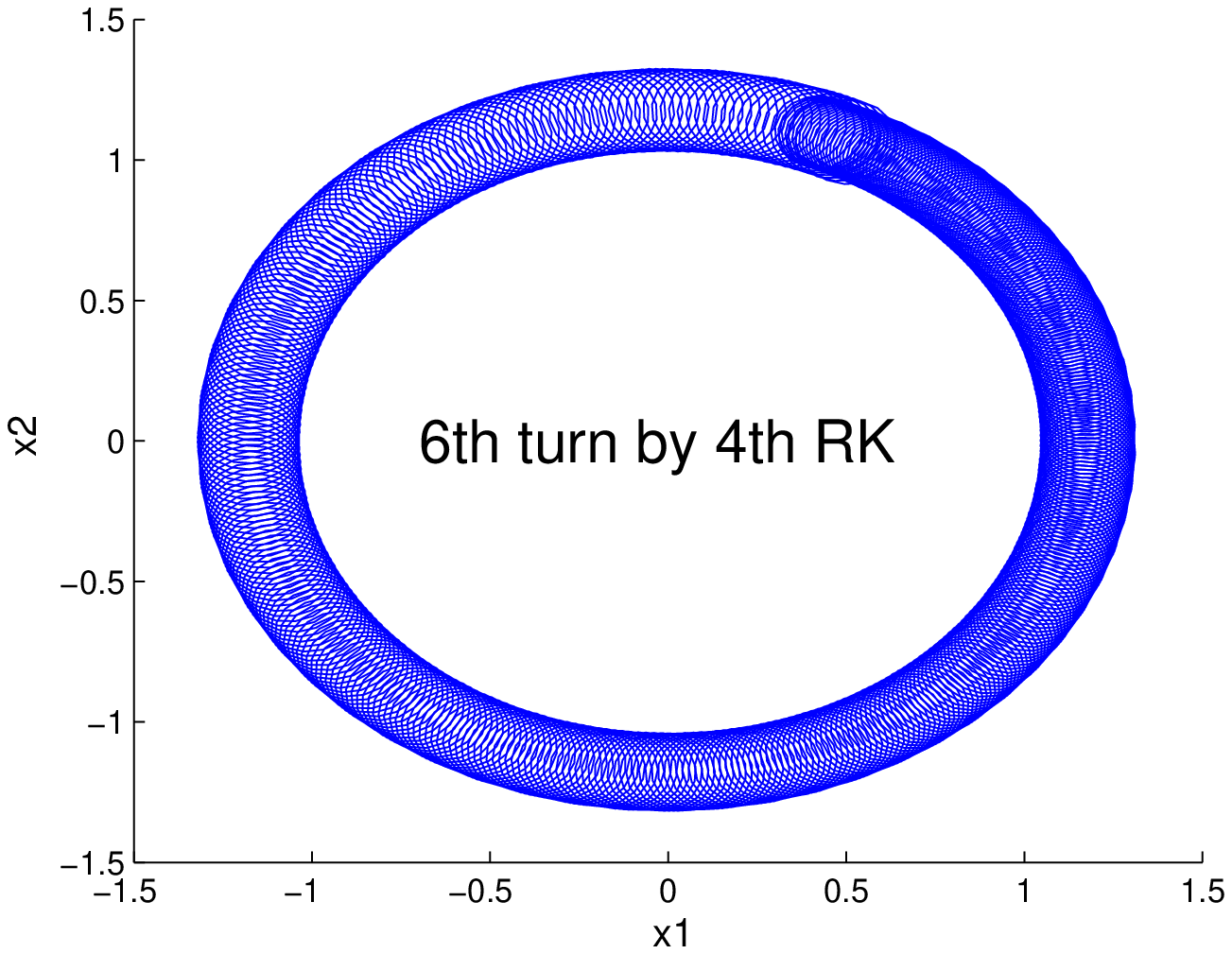}}
\subfigure[]
{\includegraphics[width=7.6cm,height=4cm]{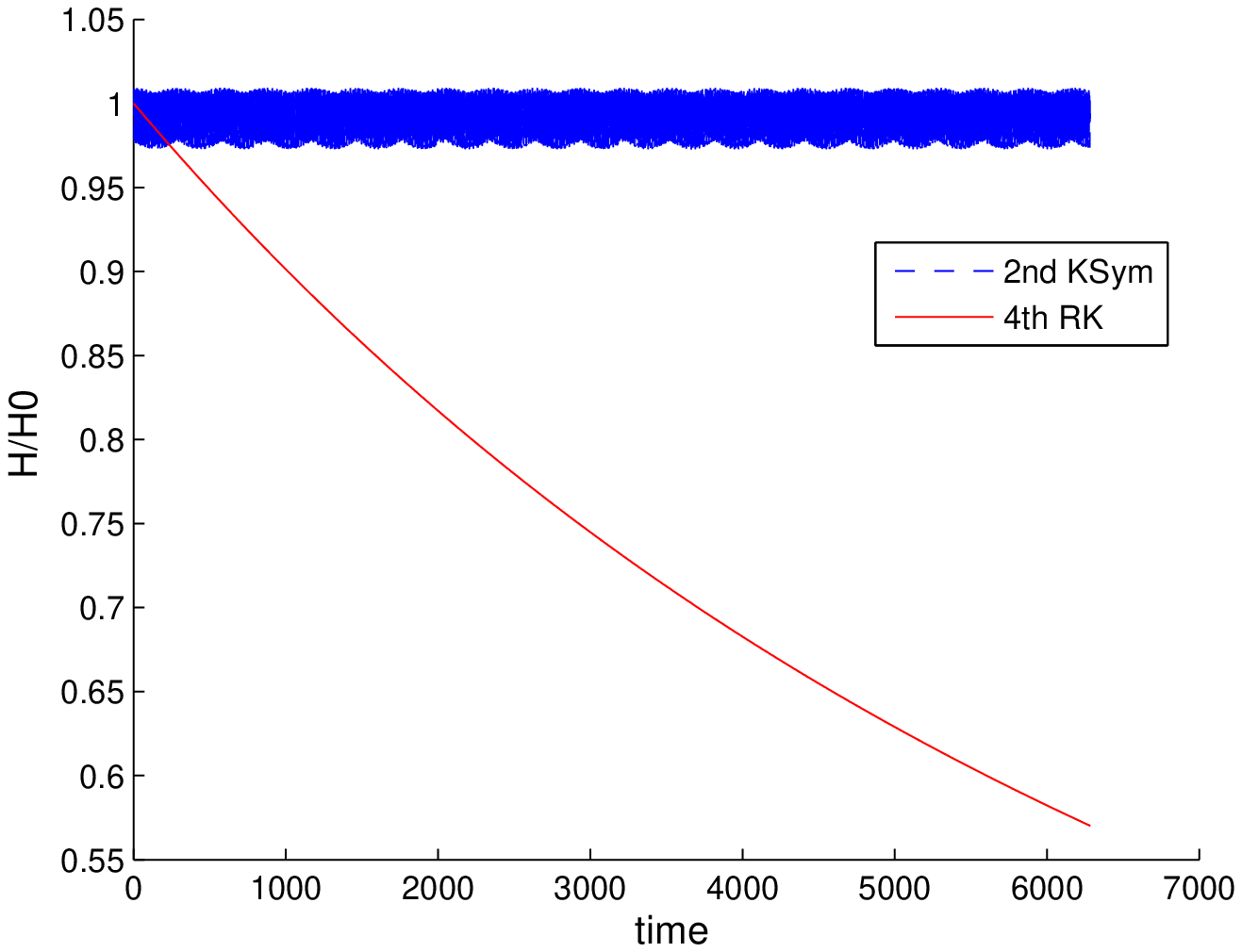}}
\subfigure[]
{\includegraphics[width=7.6cm,height=4cm]{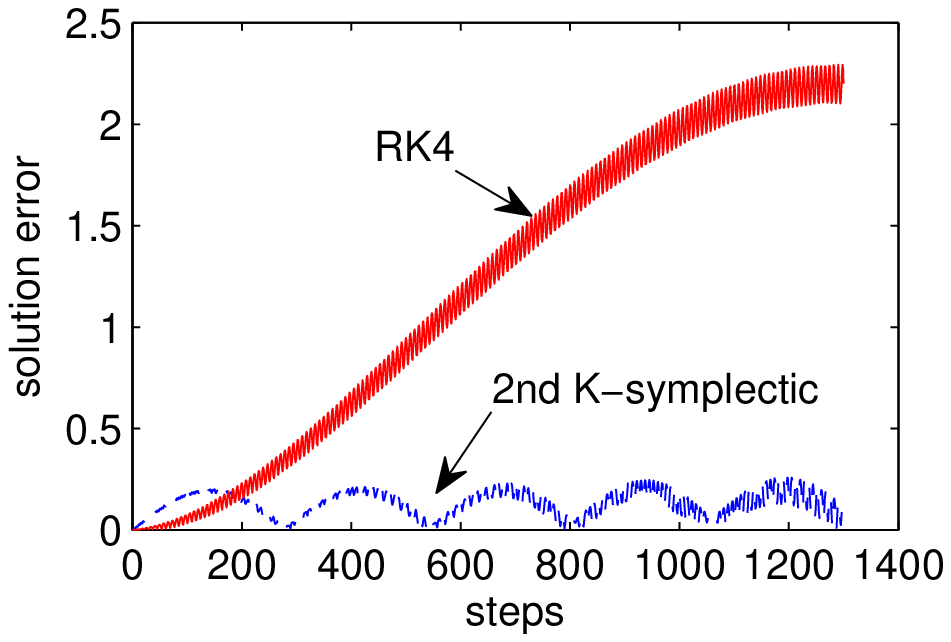}}
 \caption{(a)  Numerical orbits of the 2nd order K-symplectic method from the 190000-th step. (b) Numerical orbits of the 4th order RK method from the 190000-th step. (c) Energy error  of the two methods as a function of the computing steps till the 20000-th step. (d) Solution error as a function of the computing steps, the step size in this sub-figure is  $h=0.23\pi$.
}
\label{fig:orbBR}
\end{figure}

\begin{figure}\centering
\subfigure[]
{\includegraphics[width=7.5cm,height=5cm]{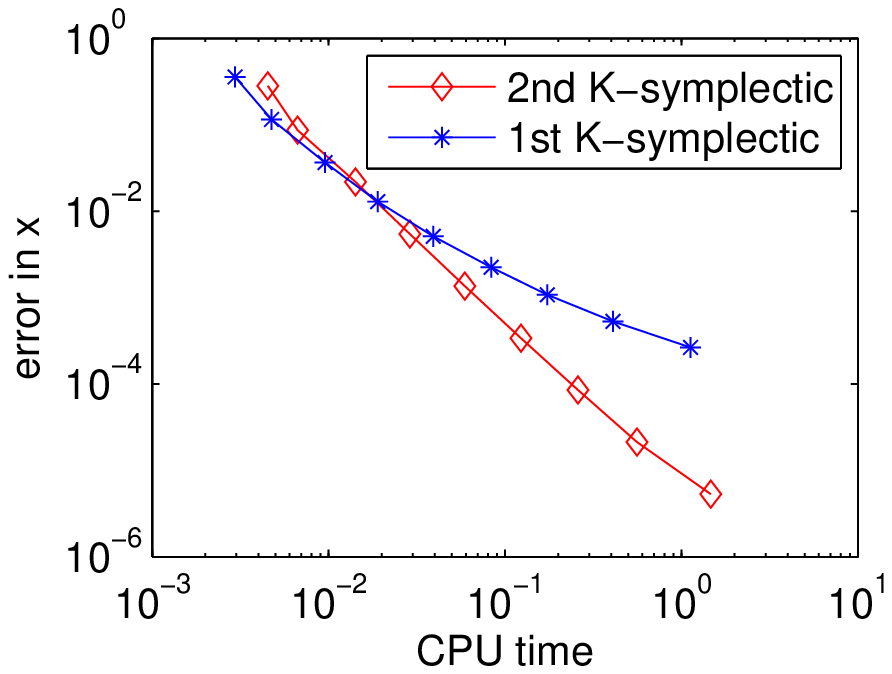}}
\subfigure[]
{\includegraphics[width=7.5cm,height=5cm]{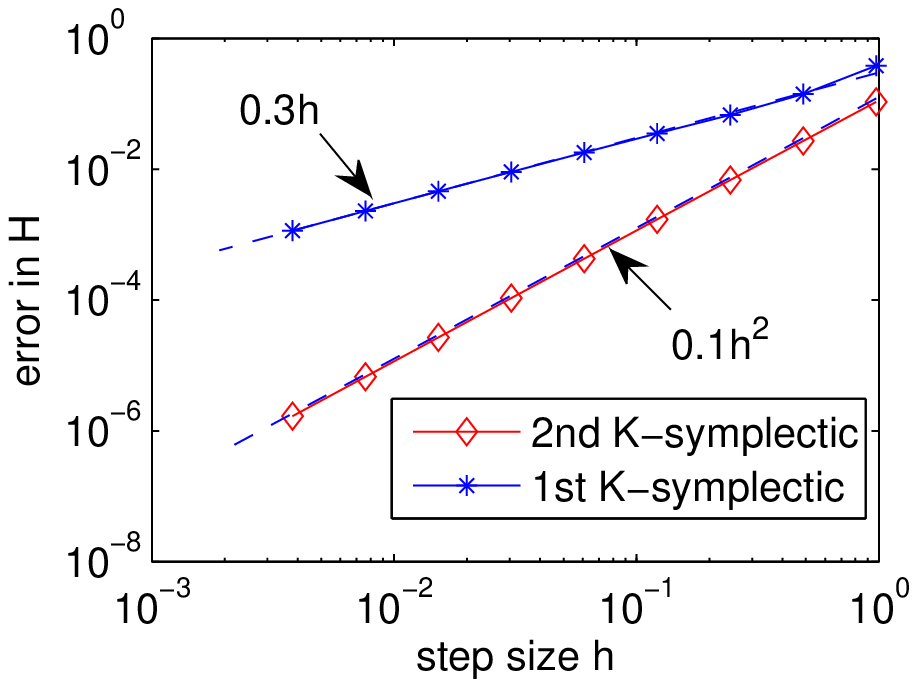}}
 \caption{(a) Solution error as a function of the computing time. (b) Error of the numericl energy as a function of the step size $h$.
}
\label{fig:ComK}
\end{figure}

In Fig.~\ref{fig:orbBR}(a) and (b), numerical orbits of the two methods from the 190000-th step are plotted. Theoretical analysis shows that the particle's orbit is a spiraling circle with constant radius, where the large circle corresponds to the  drift of the guiding center, and the small circle is the fast-scale gyromotion.  From Fig.~\ref{fig:orbBR}(a) it is observed that after long computation time
the 2nd order K-symplectic method still gives a correct orbit, while in Fig.~\ref{fig:orbBR}(b) the 4-th order RK method fails because of the dissipated  numerical gyromotion. This can be explained partly by their performances at the  energy. The obvious  energy  damping by
the 4nd order Runge-Kutta method is demonstrated in Fig. \ref{fig:orbBR}(c). On the contrary,  the second order $K$ symplectic numerical method can bound   the  energy up to error $h^2$ (see Figs.\ref{fig:orbBR}(c), \ref{fig:ComK}(b)) which  agrees with the theoretical error analysis. In Fig.\ref{fig:orbBR}(d), we show the error growth of the numerical solution along with the simulation time. The step size is chosen to be $h=0.23\pi$. It can be seen that the error caused by the Runge-kutta method is smaller at the first few steps, but grows larger than that of the $K$-symplectic method rapidly.

Next, we study the accuracy and computing amount of the $K$-symplectic methods. Fig.\ref{fig:ComK}(a) displays the solution error as a function of the computing cost. When the requirement on solution error is smaller than $1\%$, the second order method needs less computing time and is more efficient than the first order method.

\section{Conclusion}
In this paper, we have constructed numerical methods for the Lorentz force system focusing on its $K$-symplectic structure.
As there exists  a Darboux transformation such that
the $K$-symplectic structure can be turned into  the standard symplectic  structure, the error analysis for the canonical numerical methods can be generalized  to the $K$-symplectic numerical methods.
However, the Darboux transformation usually is not able to be expressed explicitly.  Thus, it is generally not easy to design the $K$-symplectic  numerical methods. For our study, we decompose successfully the Lorentz force system  as four subsystems which can be solved explicitly. The resulting numerical methods by composing the exact solutions of the four subsystems  naturally preserve the
$K$-symplectic structure. The numerical experiment shows the superior long-term performances  of the new derived numerical methods. More theoretical and  numerical analysis of the K-symplectic methods will be reported in future
publications. The splitting idea has been generalized to construct the numerical methods for the Vlasov-Maxwell equations in
\cite{heqinsun15him}.
\section*{Acknowledgments.}
This research was supported by the Fundamental Research Funds for the
Central Universities (WK2030040057), the National Natural Science Foundation
of China (11271357, 11261140328, 11305171, 11321061), by the CAS
Program for Interdisciplinary Collaboration Team, and the ITER-China Program
(2013GB111000, 2014GB124005, 2015GB111003), JSPS-NRF-NSFC
A3 Foresight Program in the field of Plasma Physics (NSFC-11261140328).

\end{document}